# Deep learning improved autofocus for motion artifact reduction and its application in quantitative susceptibility mapping

Chao Li[1,2], Jinwei Zhang[2,3], Hang Zhang[4], Jiahao Li[2,3], Pascal Spincemaille[2], Thanh D. Nguyen[2], Yi Wang[2,3]

[1]Department of Applied and Engineering Physics, Cornell University, Ithaca, NY, USA

[2]Department of Radiology, Weill Cornell Medicine, New York, NY, USA

[3]Meinig School of Biomedical Engineering, Cornell University, Ithaca, NY, USA

[4]Department of Electrical and Computer Engineering, Cornell University, Ithaca, NY, USA

**Corresponding author**:

Yi Wang, PhD

Department of Radiology

Weill Cornell Medicine

407 East 61st Street

New York, NY 10065, USA

Phone: (646) 962-2631

Email: yiwang@med.cornell.edu

**Word count:** ~2800

**ABSTRACT**

**Purpose**: To develop a pipeline for motion artifact correction in mGRE and quantitative susceptibility mapping (QSM).

**Methods**: Deep learning is integrated with autofocus to improve motion artifact suppression, which is applied QSM of patients with Parkinson's disease (PD). The estimation of affine motion parameters in the autofocus method depends on signal-to-noise ratio and lacks accuracy when data sampling occurs outside the k-space center. A deep learning strategy is employed to remove the residual motion artifacts in autofocus.

**Results**: Results obtained in simulated brain data (n =15) with reference truth show that the proposed autofocus deep learning method significantly improves the image quality of mGRE and QSM (p = 0.001 for SSIM, p < 0.0001 for PSNR and RMSE). Results from 10 PD patients with real motion artifacts in QSM have also been corrected using the proposed method and sent to an experienced radiologist for image quality evaluation, and the average image quality score has increased (p=0.0039).

**Conclusions**: The proposed method enables substantial suppression of motion artifacts in mGRE and QSM.

**INTRODUCTION**

Quantitative susceptibility mapping (QSM) is an MRI post-processing technique to obtain tissue magnetic susceptibility from the tissue field measured using complex multi-echo gradient echo (mGRE) data (1). QSM has excellent detection sensitivity and accuracy for mapping brain iron (2-9), which is valuable for assessing abnormal iron deposition in patients with Parkinson's disease (PD) (10). However, PD patients often have involuntary tremors, causing blurring and ghosting motion artifacts (11, 12).

A variety of motion compensation methods have been developed to mitigate the impact of motion in MRI, which can be broadly categorized as prospective and retrospective. Prospective methods measure organ motion (e.g., translation and rotation) prior to data acquisition by means of a MR navigator pulse (13-15) or external sensor devices such as camera (16) or magnetic field probes (17), and use the motion information to control the subsequent k-space sampling through a gating algorithm or real-time motion correction (15, 18-22). Retrospective methods estimate motion information and perform motion correction in the acquired data acquisition, which offers the advantage of no modification in data acquisition and no increase in scan time. An example of retrospective motion correction is autofocus that estimates motion parameters in each sampled echo from the motion-corrupted image data by minimizing the image entropy (23) and correct for motion artifacts using projections onto convex sets (POCS) (24, 25) or using partially-parallel-imaging-based data regeneration (26).

We propose to apply retrospective motion correction methods to existing QSM data from PD patients. The autofocus method suffers from motion estimation error when data signal-to-noise ratio (SNR) is poor, such as data away from k-space center(23). Recently, deep neural networks, including 3D conditional generative adversarial network (27) and U-Net (28), have applied to reduce motion artifacts without estimating specific motion parameters. Accordingly, we propose to combine an autofocus algorithm (23) with a deep neural network trained to learn and suppress the residual motion artifacts.

## METHODS

**Multiecho gradient echo (mGRE) signal model with motion effects**

QSM is derived from the complex mGRE signal, which at a voxel centered at position $r$ at echo time $TE$ can be expressed as

$$S(r, TE) = m(r)e^{-\frac{TE}{T2^*(r)}}e^{-ib(r)\omega_0 TE}, \qquad (1)$$

where $m(r)$ is the proton density (weighted by T1 relaxation, RF flip angle, receiver coil sensitivity, among other tissue and imaging parameters), $T_2^*$ is the effective transverse relaxation time, $\omega_0 = \gamma B_0$ is the Larmor frequency of proton in a main magnetic field $B_0$, $\gamma$ is the gyromagnetic ratio, and $b(r)$ is the total field inhomogeneity (4). After background field removal (29), the local tissue susceptibility $\chi$ can be obtained from the tissue field $b_t$ by solving a magnetic field-to-susceptibility source inverse problem,

$$b_t(r) = (d * \chi)(r) \qquad (2)$$

where $d$ is the field generated by a unit susceptibility source called the dipole kernel, and $*$ denotes the spatial convolutions.

We assume that motion occurring within the short imaging TR (approximately 50 ms or less) is negligible and therefore only consider motion between TRs. We further assume that all motion is general linear affine (which includes global translation, rotation) and that it can be described as sequence of discrete motion states, such that the only motion is the one between successive states. The motion corrupted mGRE data is then expressed as the inverse Fourier transform of the motion-corrupted k-space which is a combination of k-space lines acquired at different motion states:

$$S_m(r, TE) = IFFT\left\{\sum_i M_i \odot FFT\left[m(A_i \cdot r) e^{-\frac{TE}{T2^*(A_i \cdot r)}} e^{-id*[\chi(A_i \cdot r)]\omega_0 TE}\right]\right\}. \qquad (4)$$

Here $A_i$ is the affine matrix acting on the position $r$ at motion state $i$, $\odot$ denotes pointwise multiplication, and $M_i$ is a binary mask indicating the k-space lines acquired during motion state $i$. The phase factor in Eq. 4 is written as an affine transformation acting on the position of the susceptibility source which is then convolved with the dipole kernel.

In this work, we aimed to correct motion artifacts caused by translation and rotation, which are the primary motion components of the brain during a head MRI scan. For purely translational motion, note that $d * [\chi(A_i \cdot r)] = [d * \chi](A_i \cdot r) = b(A_i \cdot r)$, i.e., the order of the translation transformation and dipole convolution can be exchanged. While this property does not strictly hold for rotation, the intra-scan head rotation angle is relatively small (within 5°) and it is reasonable to approximate:

$$S_m(r, TE) \approx IFFT\left\{\sum_i M_i \odot FFT\left[m(A_i \cdot r) e^{-\frac{TE}{T2^*(A_i \cdot r)}} e^{-ib(A_i \cdot r)\omega_0 TE}\right]\right\}. \qquad (5)$$

This offers a feasible way to simulate the effect of translational and rotational head motion on mGRE signal. Eq.5 involves a transformation (translation and rotation) of the original image volume following with a Fourier transformation for each readout. The computation of Eq.5 is efficiently implemented in the following manner. As a rotation in the spatial domain is equivalent to a rotation of the same angle in the frequency domain, and a translation in the spatial domain is equivalent to a linear phase modulation in the frequency domain, we simulate 3D rotations by simply warping the k-space following a series of rotated readout lines and simulate translations by point-wise multiplying each readout line with a linear phase factor; an simplified demonstration is shown in Figure 1a.

## Algorithm

We propose a motion correction method comprising 1) an autofocus step to obtain an initial motion suppressed image together with the estimated motion parameters $\boldsymbol{\theta} = (\theta_t)_{t=1,\ldots,N_{TR}}$ with $N_{TR}$ the number of TRs in the acquisition; and 2) a deep learning step, where we train an end-to-end motion deep network $\Phi$ on motion-free and motion-corrupted image pairs that are generated using Eq. 5 with a range of randomly scaled versions of the image parameters $\boldsymbol{\theta}$. Accordingly, we name the proposed method as autofocusDL.

### Step 1. Autofocus

The autofocus method GradMC (23) estimates the motion parameter $\widehat{\boldsymbol{\theta}}$ which yields the sharpest image by minimizing the following cost function:

$$\widehat{\boldsymbol{\theta}} = \underset{\boldsymbol{\theta} \in \Theta}{\mathrm{argmin}}\ \phi(F^H A_{\boldsymbol{\theta}} y) + \lambda ||\nabla_t \boldsymbol{\theta}||^2 \qquad (7)$$

where $\nabla_t$ is the temporal gradient (implemented as finite differences) which imposes smoothness of the motion trajectories in time, and $\phi$ is the gradient entropy:

$$\phi(u) = H(\nabla_x u) + H(\nabla_y u)$$

where $\nabla_x$ and $\nabla_x$ are the spatial gradients and $H$ the entropy defined as

$$H(v) = -(v)^T \ln(v)$$

where $w$ is the normalized image volume in spatial domain.

$$v = \sqrt{\frac{I \odot \bar{I}}{I^H I}}$$

where $\bar{I}$ denotes the complex conjugate the image $I$, and $\odot$ denote the pointwise multiplication.

Each $\hat{\theta}_t$ in $\hat{\boldsymbol{\theta}}$ contains six parameters describing the rigid-body motion. We performed this algorithm on the first echo of our mGRE data, as it has the highest SNR. In our preliminary testing of this algorithm in simulated data for which the motion trajectories are known, we made two observations:

1. The accuracy of the estimated trajectories was dependent on the location of the data in k-space. This was also observed in (23), where it was found that the motion estimation error was relatively small for k-space data acquired near the k-space center but increased for that acquired towards the edge of k-space. The authors attributed this observation to the inverse power law drop of the signal strength as a function of the k-space radius. We also observe the same as shown in Figure S2 in the supporting information where 3 trapezoid-shaped translational motion events were simulated on a motion-free brain image and predicted using GradMC; the ability of GradMC to capture the event profiles drops towards the high-frequency regime.
2. Outside a central k-space region, the obtained trajectories roughly capture the correct timing but with an incorrect (lower) amplitude. This can be seen in Figure S2 event 3. For this large translational displacement, the algorithm approximately captures the location and the trend of the trapezoid but not its full magnitude. This is likely due to the high non-linearity of the problem and the interplay between the gradient entropy and regularization terms in the objective, causing the objective to become stuck at a local minimum when attempting to explore the full profile.

Step 2. Deep learning

Because GradMC is generally able to suppress motion artifacts originated from a central k-space region, we design here a neural network that learns to suppress the motion artifacts on data acquired with motion only at outer region of k-space. In this work, we define the central k-space region as the central 1/3 of the k-space in the phase-encoding direction, the rest of k-space will be referred to here as the k-space periphery. Next, we fine-tune a pretrained general temporal fusion U-net (see details below, Figure 2a) to suppress motion that is similar in shape as the trajectories obtaining in Step 1, except for the central k-space region, where the motion is set to zero. Specifically, we use a separate motion-free data set (25 PD patients, 20 males and 5 females, age $64.47 \pm 9.5$) acquired with the same imaging parameters and construct a training data set consisting of three simulated motion corrupted images using the trajectory from Step 1 scaled by scale 1, 1.5, and 2, respectively. For each of these trajectories, all motion parameters were set to zero for the central k-space region (Figure 2d). To limit computation time and prevent overfitting (which may erase all the other features learned by the pretrained model), fine-tuning used only 1 epoch. Finally, we applied the fine-tuned network to the mGRE volume obtained in Step 1 to suppress residual motion artifacts.

The initial supervised training of this network was based on a training dataset consisting of motion-free (see below) and simulated motion-corrupted image pairs. 5 simulated trajectories were randomly generated for each training example and were only non-zero in the k-space periphery (one example of simulated k-space trajectory is shown in Figure 1b.

**Data acquisition and preprocessing**

The human imaging studies in this work were approved by the local institutional review board and all subjects provided informed consent.

Motion-free dataset for motion artifact simulation

This dataset used for network training consists of 3D mGRE scans with negligible motion artifacts acquired from 25 PD patients on a 3T Siemens Magnetom Skyra scanner (Siemens Healthcare, Erlangen, Germany). Imaging parameters were: flip angle = 15°, first TE = 6.69 ms, TR = 44 ms, #TE = 10, ΔTE = 3.6 ms, matrix size= 260×320×256, voxel size = 0.8×0.8×1 mm$^3$. The view order involves multiple alternations of high-low-high-frequency views (in the slice encoding direction) acquired along the phase-encoding direction.

Testing datasets with real motion artifacts

The testing dataset in this study consists of 10 PD patients with varying degrees of motion artifacts selected by a neuroradiologist with 25 years of experience.

**Motion artifact simulation**

Motion-corrupted k-space data were simulated as described above, using the same motion parameters across echoes in each TR. Five motion trajectories were simulated using motion-free mGRE data for each of the 25 PD patients, resulting in a total of 125 motion-corrupted mGRE dataset, 95 for network training, 15 for validation, and 15 for testing.

Rigid body motion was simulated in the k-space periphery where autofocus estimates motion parameters poorly. As in (27), simulated motion included the 4 translational and rotational parameters corresponding to head-nodding and head shaking, which are the most common head motion in MRI scans. X/Z-rotations and X/Z-translations were randomly drawn from a Gaussian distribution with mean 0 and standard deviation 3° and 3 mm respectively, while the other 2 parameters (Y-rotation and Y-translation) were drawn from a Gaussian with mean 0 and variance 1° and 1 mm.

**Deep neural network**

We implemented a 2D bidirectional convolutional recurrent neural network (BCRNN) followed by a 2D U-Net (30) as shown in Figure 2a-c. The BCRNN serves as a temporal fusion module proposed in ref.(31, 32). In our case, a recurrent block is repeated $N_T$ times for echo sequences arranges in forward order and reverse order, where in the forward-ordered sequence $j$-th repetition is fed with $j$-th echo while in the reverse-ordered sequence $j$-th repetition is fed with $(N_T - j + 1)_{th}$ echo. The hidden features $h_i$ output by all repetitions are concatenated in the channel dimension and sent into the 2D U-Net.

**QSM reconstruction**

QSM was reconstructed for all (corrupted and corrected) mGRE volumes. Local tissue field was estimated using non-linear fitting across multi-echo phase data (33), followed by a rapid opensource minimum spanning tree algorithm (ROMEO) based phase unwrapping (34) and background field removal (29). The MEDI algorithm (35) was used to obtain the motion corrected QSM image.

**Statistical analysis**

For the simulated datasets, the motion artifact suppression performance was examined by calculating the root mean square error (RMSE), peak SNR (PSNR), and structural similarity index (SSIM) between the motion corrupted and corrected QSM images referenced to the motion-free QSM image. A paired t-test was used to assess the difference in these metrics between the motion corrupted and corrected QSM images. For the PD dataset, where the ground truth motion-free images were not available, the original and corrected images were assessed by an experienced radiologist; the image quality was graded based on the following scale: 1 = poor,

2 = diagnostic, 3 = good, and 4 = excellent. Image quality scores were compared using a Wilcoxon signed rank test. P values lower than 0.05 were assumed to indicate statistical significance.

## RESULTS

The algorithm was run successfully on all data tested. Average computation time was 2 hours including GradMC computation, network fine-tuning (data generation and training) and network inference on a NVIDIA GeForce GTX 1080 Ti GPU.

Figure 3 shows an example of mGRE and the reconstructed QSM from a representative PD subject with simulated artifacts. Compared to the ground truth, the motion artifacts were increasingly suppressed from autofocus (RMSE=43.19 for mGRE and RMSE=0.0122 ppm for QSM) to autofocusDL (RMSE = 36.54 for mGRE and RMSE = 0.0110 ppm for QSM ).

Figure 4 shows three examples from the in vivo PD datasets with actual motion artifacts. The proposed algorithm was effective in suppressing motion artifacts, especially the ghosting, leading to better image quality.

Figure 5a-c shows boxplots comparing the SSIM, PSNR and RMSE of the motion-corrupted and motion-corrected QSMs for the validation set including 15 simulated examples. Compared to the motion-corrupted images, the motion-corrected images achieved statistically significant higher SSIM ($0.966 \pm 0.019$ versus $0.940 \pm 0.021$, $p = 0.001$) and PSNR ($39.237 \pm 0.606$ versus $37.072 \pm 0.823$, $p < 0.0001$) as well as lower RMSE ($0.0109 \pm 0.0008$ versus $0.0141 \pm 0.014$, $p < 0.0001$).

Figure 5d compares the QSM image quality scores provided by the expert reader for the 10 in vivo motion-corrupted 10 PD cases. Without motion correction, 8 cases were rated as poor with the lowest image quality score of 1. Motion correction significantly improved the average image quality score from $1.20 \pm 0.42$ to $2.40 \pm 0.69$ ($p = 0.0039$).

## DISCUSSION

Our preliminary results indicate that the proposed autofocusDL method can improve upon autofocus for retrospectively correcting motion artifacts in QSM. Our results include both simulation data where a motion-free ground truth was available, and clinical PD patient data where involuntary motion occurs commonly. The autofocus method suffers residual motion artifacts, as motion estimation error increases as k-space data SNR decreases. The residual motion artifacts can be further reduced using deep learning.

Motion artifacts tend to exhibit various patterns depending on the k-space position, duration and magnitude of motion. While motion occurring at low spatial frequencies causes blurring, motion occurring when the high spatial frequencies are sampled causes ringing artifacts(11). Because the space of all possible motion trajectories is very large, motion correction by deep learning alone, typically using an end-to-end deep neural network (DNN) trained on a limited dataset, can be suboptimal. DNN with limited capacity averages the artifacts, and this averaging can introduce unwanted blurriness into the output images. Furthermore, simulating motion in the low frequency of k-space in training data may cause a global misalignment between the motion-corrupted image and the ground truth, leading to more blurriness in the network output. In the proposed autofocusDL method, the motion during the acquisition of the central k-space region is estimated and corrected for using an autofocus method, and DNN is trained only on a subspace of motion trajectories where motion occurs at k-space periphery and motion blurring artifact is small. With this heuristic explanation, the autofocus step in autofocusDL can track and correct for motion at the k-space center and the deep learning step in autofocusDL can compensate for motion at the k-space periphery, and correspondingly, autofucusDL can reduce ringing artifacts without increasing blurring artifact.

Unlike other MRI techniques where magnitude image matters for diagnosis purpose, QSM uses the phase image information of mGRE data to calculate susceptibility distribution. Strictly speaking, this phase image does not undergo the same affine motion transformation as the magnitude image but instead is equal to the dipole convolution of the affine transformed susceptibility distribution, leading to the well-known observation that the phase (or field) is orientation dependent (36). To simplify the calculation, in this work we assumed that patient motion falls within a small range and therefore the order of affine transformation and dipole convolution can be interchanged as an approximation. This avoids the issues related to dipole convolution, background field and brain erosion in the forward model when generating the training data for the proposed deep learning method. By warping the k-space to simulate rotated readout lines and modulating the phase of the k-space to simulate translations for each echo, a motion corrupted mGRE volume is obtained by inverse Fourier transform. From this, a simulated motion corrupted QSM is computed using regular dipole inversion.

The proposed autofocusDL method has several limitations. First, the autofocus method is computationally expensive. The current implementation could be translated into clinical practice by selectively applying it to those patients with the most severe motion artifacts, which in our group was 5% of the PD patients. Second, there may be residual motion still in the k-space center, as well as in the k-space periphery. Further development for motion correction is needed, such as combining with prospectively acquired or real-time navigator information(15).

## CONCLUSION

The proposed autofocusDL algorithm for retrospective motion correction combines an autofocus method with deep learning and improves upon autofocus in reducing motion artifacts. AutofocusDL improved QSM image quality from non-diagnostic to diagnostic in PD patients.

# FIGURES

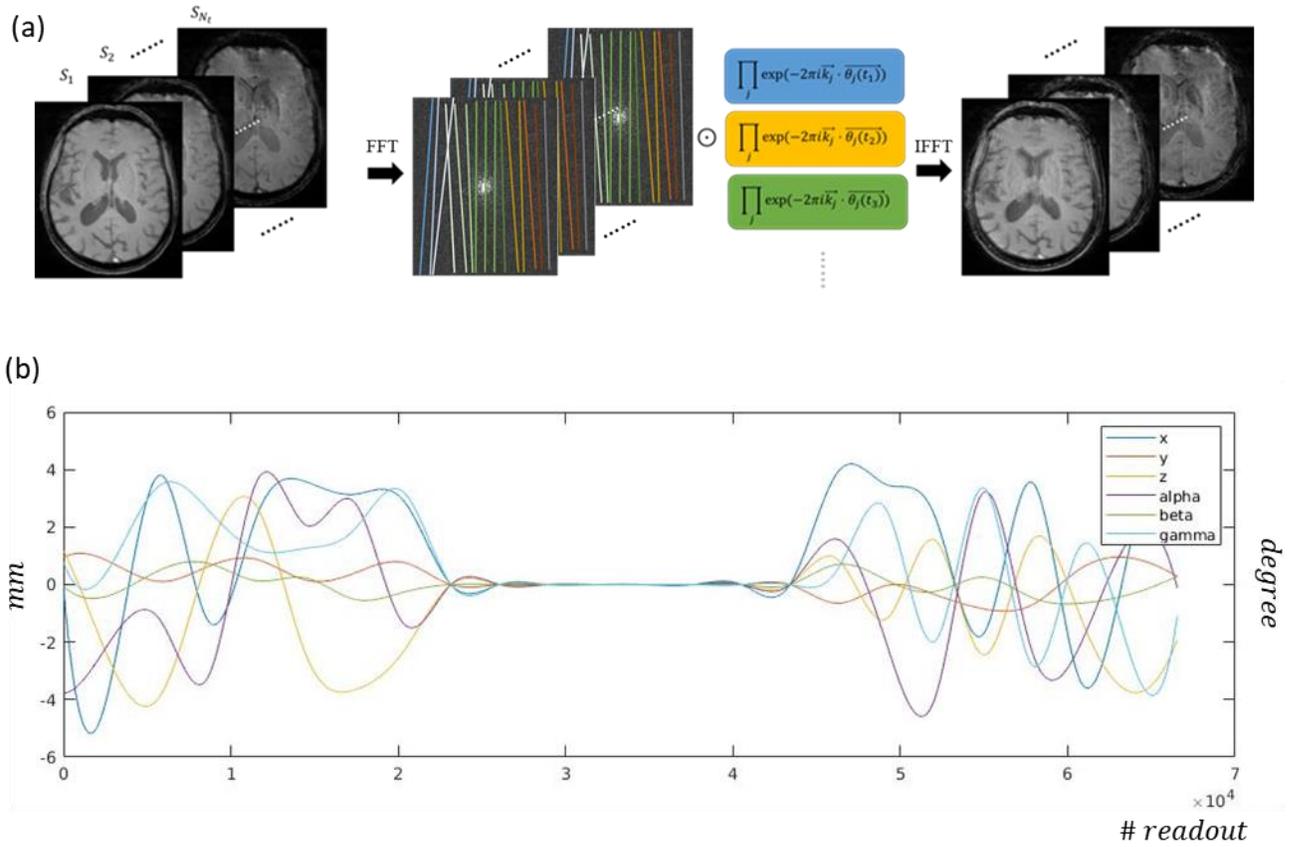

**Figure 1**. (a) Workflow for simulating motion-corrupted images. The set of readout-lines with different color corresponds to a different motion state, assuming negligible motion between the echoes within one TR. (b) An example of simulated motion trajectories with motion only in the k-space periphery.

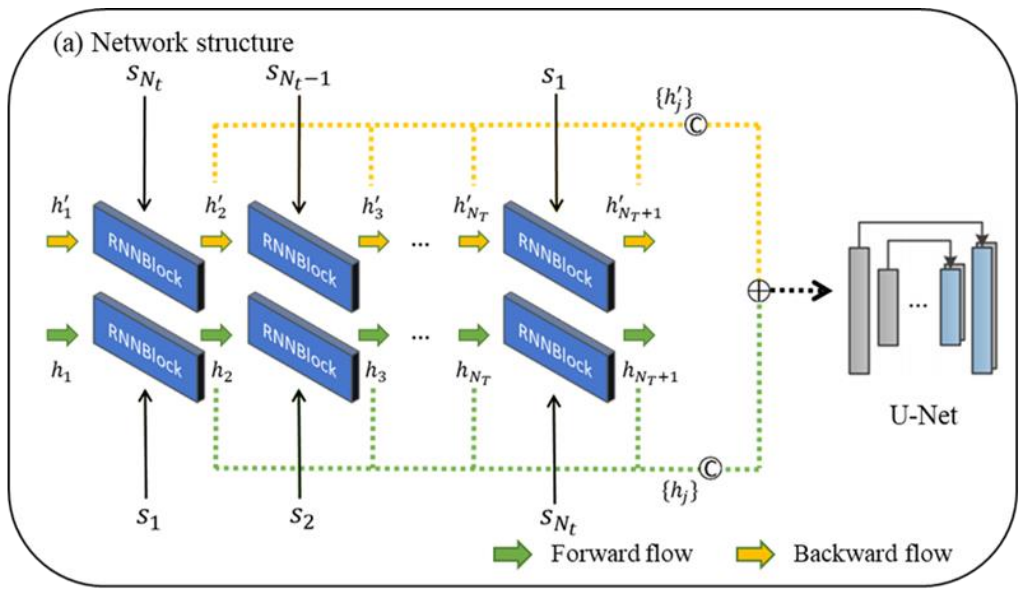

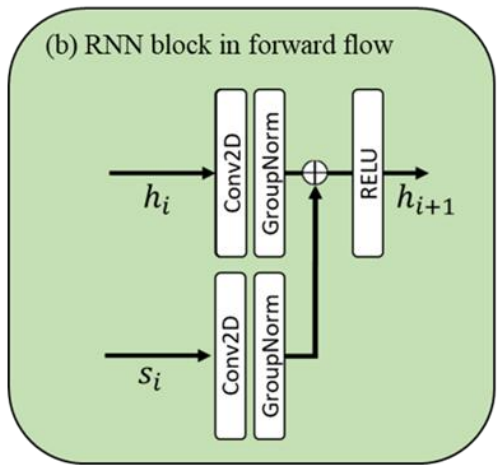
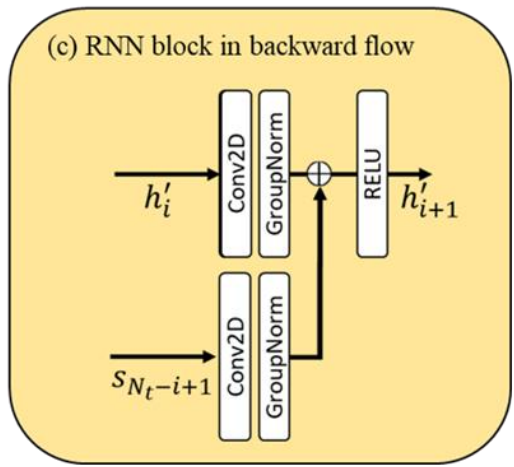

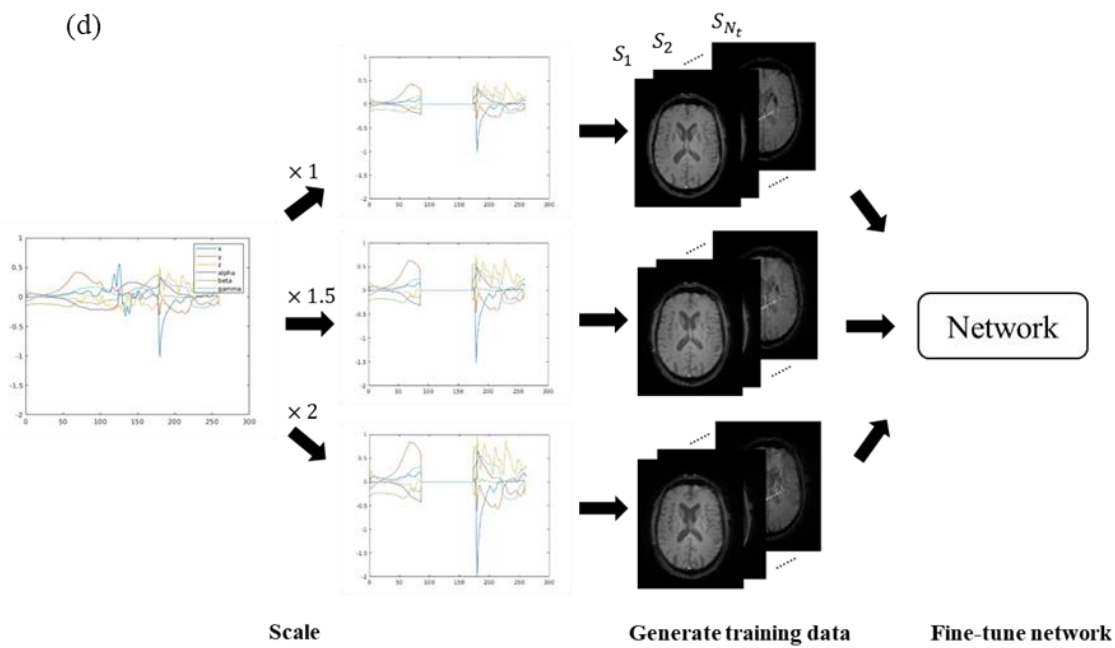

**Figure 2.** Network and fine-tuning: (a) the forward process of the network: The motion corrupted image is passed through a 2D BCRNN that fuses the features from all echoes $\{s_j | j = 1 \ldots N_T\}$. The hidden features from al repetitions $\{h_j\}$ are concatenated together and sent to a 2D UNet. (b) and (c) show the internal structures of RNN blocks for the forward flow and backward flow respectively. (d) Flowchart for fine-tuning the network with subject-specific motion trajectories. The estimated trajectory obtained by GradMC is scaled to generate training data for network fine-tuning.

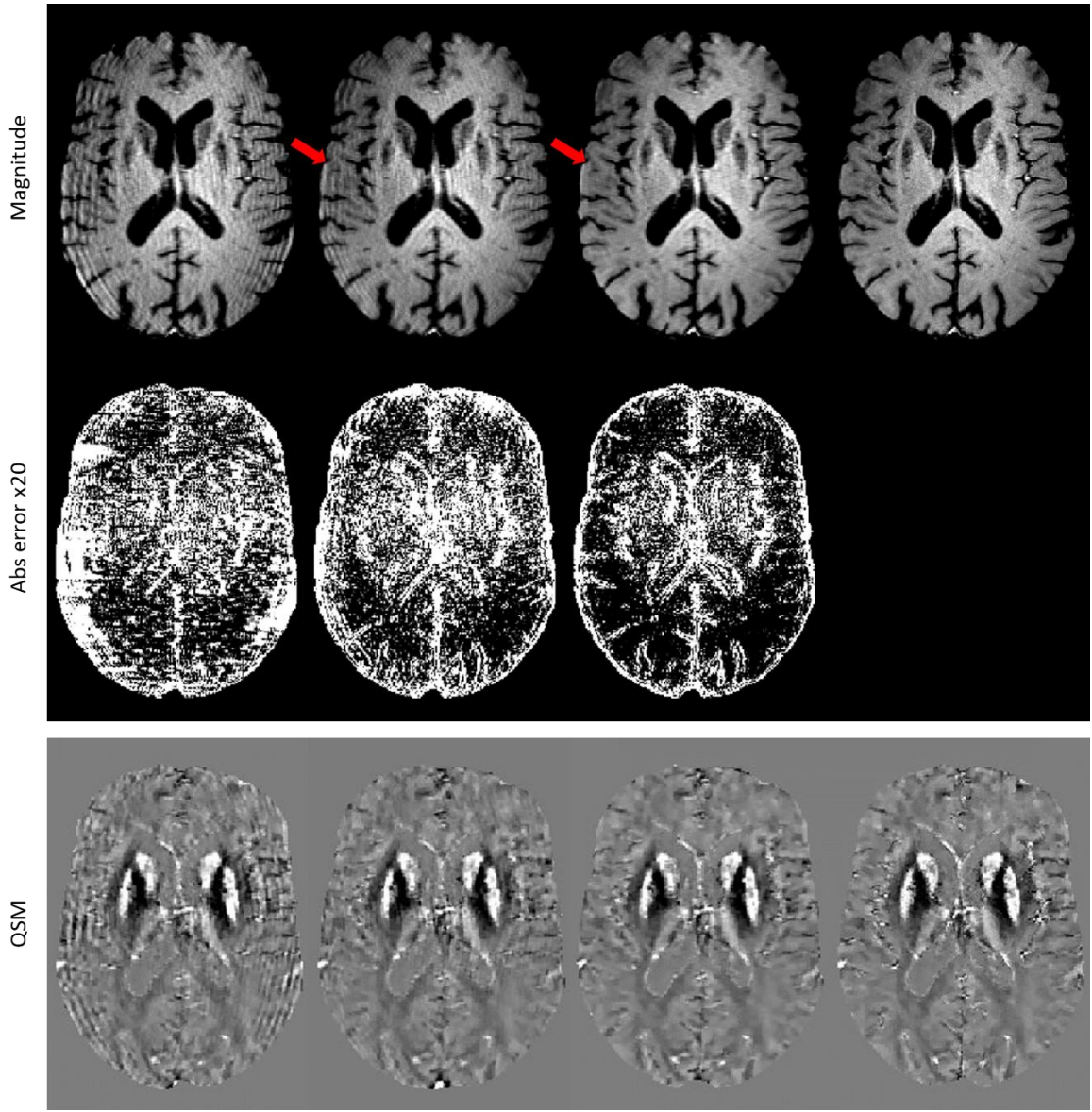

**Figure 3.** An example from a simulated dataset showing the first echo of the MGRE data (top), error map (middle) and the corresponding QSM (bottom). Columns depict the motion-corrupted image, the autofocus-corrected image, the result from the proposed autofocusDL method, and the motion-free reference. These images demonstrate an effective removal of ghosting artifacts at the cost of slight blurring in the corrected images.

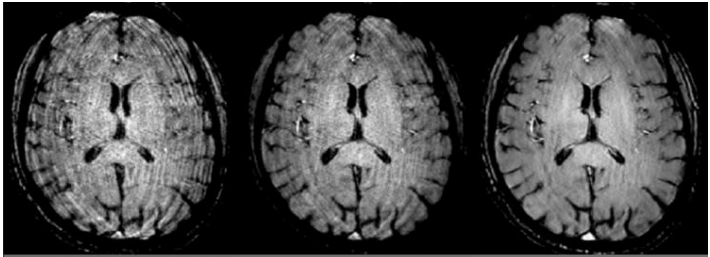
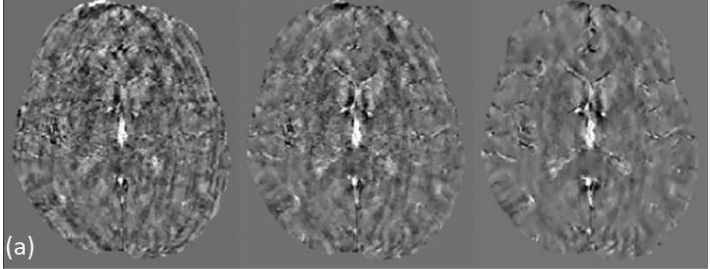
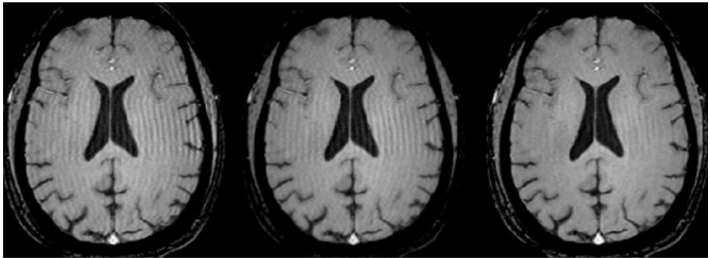
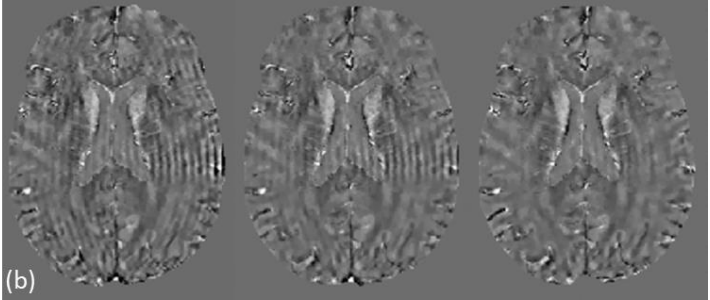
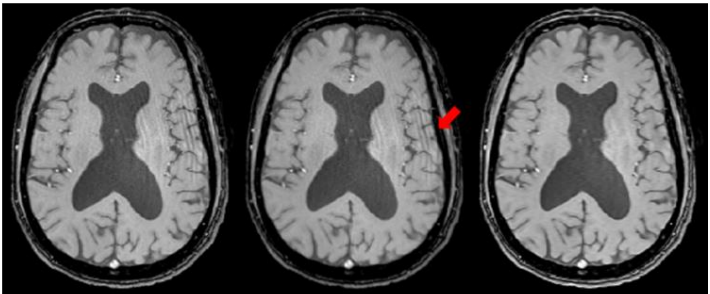
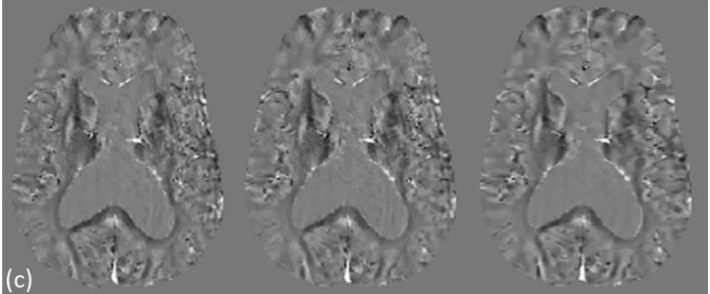

**Figure 4.** Three in vivo examples with severe (a), moderate (b) and light (c) motion artifacts. Displayed are mGRE first-echo (top) and corresponding QSM (bottom) from the original uncorrected data, autofocus corrected data, and autofocusDL corrected data.

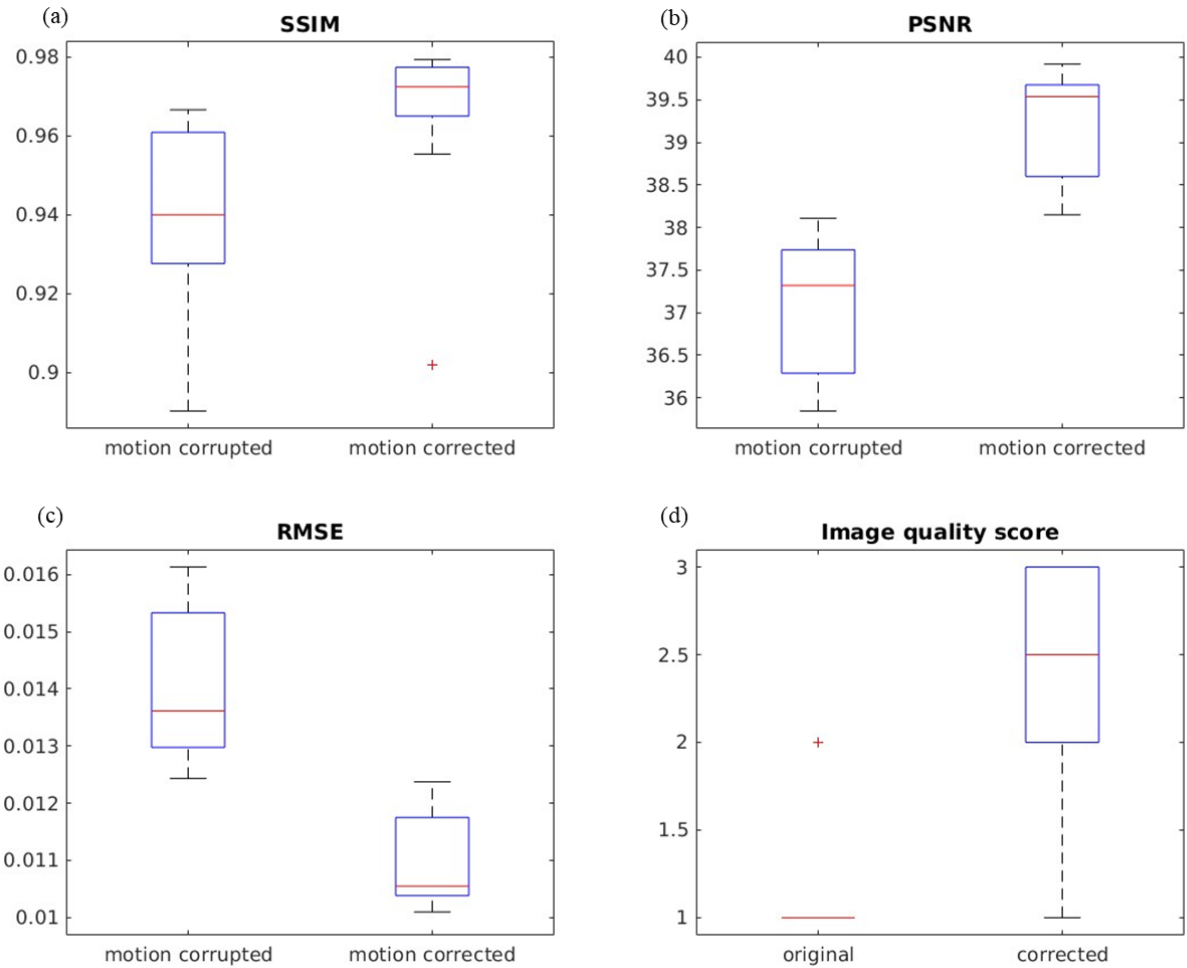

**Figure 5.** Comparison of motion-corrupted and motion-corrected images with regards to a) SSIM, b) PSNR, and c) RMSE obtained in the simulated dataset (using the motion-free image as reference), and d) QSM image quality score obtained in the in vivo PD dataset. For each box plot, the central, top, and bottom mark indicate the median, the 25[th] percentile, and the 75th percentile, respectively. All differences were statistically significant ($p < 0.0001$).